\documentclass{article}

\usepackage{arxiv}

\usepackage[utf8]{inputenc} 
\usepackage[T1]{fontenc}    
\usepackage{hyperref}       
\usepackage{url}            
\usepackage{booktabs}       
\usepackage{amsfonts}       
\usepackage{nicefrac}       
\usepackage{microtype}      
\usepackage{lipsum}
\usepackage{graphicx}

\usepackage{algpseudocode}
\usepackage{mathtools}
\usepackage{amsmath}
\usepackage{mathtools}
\usepackage{amssymb}
\usepackage{bm}
\usepackage{algorithm}
\usepackage{algpseudocode}
\usepackage{xcolor}

\usepackage{graphicx} 
\usepackage{subcaption} 

\graphicspath{ {./images/} }

\title{Bayesian defective Marshall-Olkin Gompertz model: an integrated approach to identifying cure fraction}

\author{
  Dionisio Alves-Neto \\
  Institute of Mathematics and Computer Sciences\\
  University of São Paulo\\
  São Carlos, São Paulo, Brazil \\
  \texttt{dionisioneto@usp.br}
  \and
  \textbf{Vera Tomazella} \\
  Department of Statistics\\
  Federal University of São Carlos\\
  São Carlos, São Paulo, Brazil \\
  \texttt{vera@ufscar.br}
  \and
  \textbf{Adriano Suzuki} \\
  Institute of Mathematics and Computer Sciences\\
  University of São Paulo \\
  São Carlos, São Paulo, Brazil \\
  \texttt{suzuki@icmc.usp.br}
  \and
  \textbf{Danilo Alvares} \\
  MRC Biostatistics Unit\\
  University of Cambridge\\
  Cambridge, United Kingdom\\
  \texttt{danilo.alvares@mrc-bsu.cam.ac.uk}
}

\begin{document}
\maketitle
\begin{abstract}
Regression models have a substantial impact on interpretation of treatments, genetic characteristics and other potential risk factors in survival analysis. In many applications, the description of censoring and survival curve reveals the presence of cure fraction on data, which leads to alternative modeling. The most common approach to introduce covariates under a parameter estimation is the cure rate model and its variations, although the use of defective distributions have introduced a more parsimonious and integrated approach. Defective distributions are given by a density function whose integration is not one after changing the domain of one of the parameters, making them appropriate for survival curves with an evident plateau. In this work, we introduce a new Bayesian defective regression model for long-term survival outcomes using the Marshall-Olkin Gompertz distribution. The estimation process is under the Bayesian paradigm. We evaluate the asymptotic properties of our proposal under the vague prior scheme in Monte Carlo studies. We present a motivating real-world application using data from patients diagnosed with testicular cancer in São Paulo, Brazil, in which long-term survivors were identified. Scenarios of cure with uncertainty estimates via credible intervals are provided to evaluate characteristics such as risk age, presence of treatment, and cancer stage. 

\end{abstract}


\section{Introduction}
\label{intro}

In survival analysis, the focus turn into descriptions and inferences about the time until the
occurrence of a specific event. This event, often referred as failure time, is a random variable defined on the non-negative real numbers. The fundamental assumption about the time-to-event data is that all observed individuals will eventually experience the event of interest, i.e, all individuals are assumed to experience it at some point event if the extended follow-up does not capture the failure moment\cite{klein2014handbook}. However, with advancements in medical treatments and pharmaceuticals, a significant portion of observed individuals tend to exhibit a cure for death or recurrence of a disease \cite{maller1996survival}. \cite{lin2008cure} also discuss about this phenomena in reliability studies, where similarly situations arises when repair designs are incorporated into equipment to delay the failure time. Thus, assuming that the event of interest will occur for all sample units, despite the presence of a cure factor, leads to invalid inferences. Therefore, an alternative approach to traditional probabilistic models is necessary. 

The first methodology introduced in the literature comprises the standard mixture models of \cite{boag1949maximum} and \cite{berkson1952survival}. In this structure, the population is assumed to be divided into two groups: cured and uncured individuals. This class of models adjusts the survival function by adding a new parameter ($p$) to capture the fraction of cured individuals, as shown by $S(t) = p + (1-p) S_{0}(t)$, where $S_{0}(t)$ is a proper survival function for uncured observations. As $t \rightarrow \infty$, the survival function 
 $S(t)$ converges to $p$.  Additionally, there are variations of promotion time models in which the number of competing risks follows a Poisson distribution to represent the number of causes which promotes the event of interest\cite{tsodikov1996stochastic}, and an even more unified version where the number of competing causes follows a Negative Binomial distribution \cite{rodrigues2009unification}. 

 A recent advancement in modeling data with a cure fraction lies in the study of defective models \cite{balka2009review}. This class is defined by a proper probability distribution that, upon modifying the natural parameter space of one of its parameters, becomes improper as it no longer integrates to one \cite{scudilio2019defective}. In practical scenarios, a defective model detects the presence of a cure fraction in the observations when one of its parameter takes a value outside the corresponding original parameter space. Otherwise, parameter estimation proceeds conventionally. In this regard, this framework provides a comprehensive approach to data with or without long-term survival times. If the model assumes a defective form, the survival function decreases to a constant value $p$, indicating the fraction of cured individuals in the observed data \cite{scudilio2019defective}. This can be estimated through a derived estimand ($i.e$., a function of the estimated parameters). Defective models offer some advantages in comparison with other parameter approaches, such as standard mixture cure models: (1) no special specification is required to account for the presence of long-term survivors, which facilitates the estimation process by reducing the number of parameters; (2) the inferential procedure itself can indicate the presence of immune individuals when a parameter takes values outside the conventional parameter space; and (3) the cure fraction is a function of the estimated parameters and can be easily computed as the limit of the non-proper survival function. However, the class of defective distributions is limited to parametric options allowed to present the improper behavior. There are some famous alternatives such as the Gompertz distribution, the inverse Gaussian distribution and the Dagum distribution. 

In the work of \cite{rocha2016newtwoMO}, two new flexible defective distributions within the Marshall-Olkin family are introduced. The Marshall-Olkin family provides a framework for creating new probability distributions based on existing ones with the aim of adding more flexibility in statistial modeling. \cite{rocha2016newtwoMO} demonstrates that, by employing the defective versions of the Gompertz and inverse Gaussian distributions, the models within the Marshall-Olkin family also become defective. Mathematical properties of the Marshall-Olkin family are present in depth in the paper of \cite{barreto2013general}. Under the frequentist approach, \cite{rocha2016newtwoMO} presented simulation studies to evaluate the properties of maximum likelihood estimation and conducted model comparison experiments to demonstrate the ability to make accurate inferences for the Marshall-Olkin Gompertz distribution in comparison to the conventional defective Gompertz distribution. Three applications to different datasets were provided to showcase the practical performance of the defective model compared to the mixture cure models for these two new distributions. 

In the presence of a cure fraction, it is of particular interest to assess how treatments, genetic characteristics, and other endogenous factors affect patient immunity. Nevertheless, the influence of such covariates has not yet been investigated for defective distributions derived from the Marshall–Olkin family. The importance of covariates was explored by \cite{martinez2017defective}, who introduced the parametric regression under frequentist and Bayesian inferences for the defective generalized Gompertz model. An approach for interval-censored survival data with immune subjects for Gompertz and inverse Gaussian defective distributions is discussed by \cite{calsavara2019defective}. Therefore, the objective of this paper is to propose the Marshall-Olkin Gompertz distribution as alternatives to comprehensive models for estimating the parameter survival function in the presence of a cured fraction in the population and how clinical profiles can impact the probability of cure. In addition, this paper introduces the first approach under Bayesian paradigm for the defective Marshall-Olkin Gompertz model proposed by \cite{rocha2016newtwoMO}.

\section{Methods}
\label{sec:Methods}

\subsection{The Marshall-Olkin Gompertz cure rate regression model}

The Marshall–Olkin Gompertz distribution (MOGD) is an extension of the Gompertz distribution, which is widely employed in lifetime data analysis, particularly in contexts where an exponential-type risk behavior is known. Introduced by \cite{rocha2016newtwoMO}, this distribution provides a new alternative in the field of defective modeling and flexibility due to a new parameter added.

The MOGD is defined by three parameters and models the failure time conditional to $\alpha > 0$ (shape), $\mu > 0$ (scale), and $\Upsilon > 0$ (shape). The probability density function, the survival function, and the hazard function are, respectively, given by:
\begin{equation}
     f_{MO}(t ; \alpha, \mu, \Upsilon) = \dfrac{\Upsilon\mu \exp(\alpha t) \exp\left(-\dfrac{\mu}{\alpha} \, [\exp(\alpha t) - 1]\right)}{\left[ 1 - (1 - \Upsilon) \exp\left(-\dfrac{\mu}{\alpha} \, [\exp(\alpha t) - 1]\right)) \right]^2},
\label{eq:den_MO_gompertz}
\end{equation}

\begin{equation}
    S_{MO}(t ; \alpha, \mu, \Upsilon) = \dfrac{\Upsilon \exp\left(-\dfrac{\mu}{\alpha} \, [\exp(\alpha t) - 1]\right)}{1 - (1- \Upsilon) \exp\left(-\dfrac{\mu}{\alpha} \, [\exp(\alpha t) - 1]\right)},
    \label{eq:surv_MO_gompertz}
\end{equation}

\begin{equation}
    h_{MO}(t ; \alpha, \mu, \Upsilon) = \dfrac{\Upsilon \mu \exp(\alpha t)}{1 - (1- \Upsilon) \exp\left(-\dfrac{\mu}{\alpha} \, [\exp(\alpha t) - 1]\right)}.
    \label{eq:hazard_MO_gompertz}
\end{equation}

The MOGD has some particular cases included: ($i$) $\Upsilon = 1$, the Gompertz distribution; ($ii$) the Marshall-Olkin Exponential distribution, when $\mu=1$; and ($iii$) the exponential distribution when $\Upsilon = 1$ and $\mu=1$.

The MOGD is an interesting parametric option when immune individuals are considered in the analysis. This distribution can present the defective behavior (i.e, the cumulative distribution does not integrate 1) when the parameter space of one of its parameters is expanded. As proved by \cite{rocha2016newtwoMO}, considering $\alpha \in \mathbb{R}$, the MOGD becomes improper when $\alpha < 0$ and the cure proportion detected can be computed as  
\begin{align}
    p &= \lim_{t \rightarrow \infty} S_{MO}(t ; \alpha, \mu, \Upsilon) 
    = \lim_{t \rightarrow \infty} 
    \frac{\Upsilon \,\exp\!\left(-\frac{\mu}{\alpha}\left(\exp(\alpha t) - 1\right)\right)}
         {1 - (1-\Upsilon)\,\exp\!\left(-\frac{\mu}{\alpha}\left(\exp(\alpha t) - 1\right)\right)} \\[6pt]
    &= \frac{\Upsilon \,\lim_{t \rightarrow \infty} \exp\!\left(-\frac{\mu}{\alpha}\left(\exp(\alpha t) - 1\right)\right)}
            {1 - (1-\Upsilon)\,\lim_{t \rightarrow \infty} \exp\!\left(-\frac{\mu}{\alpha}\left(\exp(\alpha t) - 1\right)\right)} 
     = \frac{\Upsilon\, \exp(\frac{\mu}{\alpha})}{1 - (1-\Upsilon)\, \exp(\frac{\mu}{\alpha})}.
\end{align}

It means once the values of $\alpha$, $\mu$ and $\Upsilon$ are estimated, the proportion of cure in the population can be easily obtained as a function of the parameters, avoiding a special structure that may burden the inferential process and offering also an integrated framework. When $\alpha > 0$, the cure fraction ($p$) is equal to zero.

To model the distribution for clinical data, we include the covariate effects ($\boldsymbol{w}_{i\alpha}$) in the analysis to the shape parameter $\alpha$, which can be either positive (no cure fraction in data) or negative (presence of cure fraction in data), using the identity link function, that is,
\begin{equation*}
    \alpha(\boldsymbol{w}_{i\alpha}) = \boldsymbol{w}_{i\alpha}^{\top} \boldsymbol{\psi}_{\alpha},
\end{equation*}

\noindent where $\boldsymbol{w}_{i\alpha}^{\top} = (1, w_{\alpha i 1},  w_{\alpha i 2}, \dots,  w_{\alpha i q_{\alpha}})$ is a vector of observations from $q_{\alpha}$ independent variables for the $i$-th observation and $\boldsymbol{\psi}_{\alpha} = (\psi_{\alpha 0}, \psi_{\alpha 1}, \dots, \psi_{\alpha q_{\alpha}})^{\top}$ is the vector of their corresponding regression coefficients. Similarly, covariate information can also be provided to the scale parameter $\mu > 0$, because of its parameter domain, the logarithm link function is used,
\begin{equation*}
    \mu(\boldsymbol{w}_{i\mu}) = \exp(\boldsymbol{w}_{i\mu}^{\top} \boldsymbol{\psi}_{\mu}),
\end{equation*}

\noindent where  $\boldsymbol{w}_{i\mu}^{\top} = (1, w_{\mu i 1},  w_{\mu i 2}, \dots,  w_{\mu i q_{\mu}})$ is a vector of observations from $q_{\mu}$ independent variables for the $i$-th observation and $\boldsymbol{\psi}_{\mu} = (\psi_{\mu 0}, \psi_{\mu 1}, \dots, \psi_{\mu q_{\mu}})^{\top}$ is the vector of their corresponding regression coefficients. The covariates information can be the same for the scale ($\alpha$) and shape ($\mu$) parameters.

To identify the particular cases and repect the principle of parcimony, we opted to avoid adding regression coefficients in the parameter $\Upsilon$, which aggregates flexibility in the distribution. Based in this regression definitions, if the cure fraction is detected through the shape parameter, the $i-$th probability of cure is given by

\begin{equation*}
    p_{i}(\boldsymbol{w}_{i\alpha}, \boldsymbol{w}_{i\mu}) = \dfrac{\Upsilon \exp\left(\dfrac{\exp(\boldsymbol{w}_{i\mu}^{\top} \boldsymbol{\psi}_{\mu})}{\boldsymbol{w}_{i\alpha}^{\top} \boldsymbol{\psi}_{\alpha}}\right)}{ 1 - (1-\Upsilon) \exp\left(\dfrac{\exp(\boldsymbol{w}_{i\mu}^{\top} \boldsymbol{\psi}_{\mu})}{\boldsymbol{w}_{i\alpha}^{\top} \boldsymbol{\psi}_{\alpha}}\right)}, i = 1, 2, \dots, n.
\end{equation*}

Additionally, other link functions and regression structures can be proposed, but it is out of the scope of this study. The interpretation of the cure is better explained in terms of categorical covariates, where categories in $\boldsymbol{w}_{i\alpha}$ can detect cure in different groups, where the individual effects expressed in the regression coefficient. The clinical specialist can analyze other categorical variables in $\boldsymbol{w}_{i\mu}$, where bigger values in the linear combination indicates the decease in the probability of cure while  lower values the increase in the probability of cure.

\subsection{Inference}

Supposing a survival experiment under right-censoring mechanism of $n$ independent individuals with observed dataset $\mathcal{D} = \left\{T_i, \delta_i, \boldsymbol{w}_{i\alpha}, \boldsymbol{w}_{i\mu} ; i = 1, 2, \dots, n\right\}$, where $T_i$ is the survival time registered, $\delta_i$ is the event indicator, and  $( \boldsymbol{w}_{i\alpha}, \boldsymbol{w}_{i\mu})$ are covariate data for the $i-$th observation. The vector of parameter is $\boldsymbol{\vartheta} = (\boldsymbol{\psi}_{\alpha}, \boldsymbol{\psi}_{\mu}, \Upsilon)$. The likelihood function for non-informative right censored data is
\begin{align}
    \pi(\mathcal{D} \mid \boldsymbol{\vartheta}) &= \prod_{i=1}^{n} h_{MO}(t_{i} ;\boldsymbol{\vartheta})^{\delta_i} \times S_{MO}(t_{i} ;\boldsymbol{\vartheta}) \nonumber \\
    &= \prod_{i=1}^{n} \left\{\dfrac{\Upsilon \exp(\boldsymbol{w}_{i\mu}^{\top} \boldsymbol{\psi}_{\mu}) \exp(\boldsymbol{w}_{i\alpha}^{\top} \boldsymbol{\psi}_{\alpha} t_{i})}{1 - (1 - \Upsilon) \exp\left(-\dfrac{\exp(\boldsymbol{w}_{i\mu}^{\top} \boldsymbol{\psi}_{\mu})}{\boldsymbol{w}_{i\alpha}^{\top} \boldsymbol{\psi}_{\alpha}} \, [\exp(\boldsymbol{w}_{i\alpha}^{\top} \boldsymbol{\psi}_{\alpha} t_{i}) - 1]\right) }\right\}^{\delta_{i}} \nonumber \\
    & \quad \qquad \times \left\{\dfrac{\Upsilon \exp\left(-\dfrac{\exp(\boldsymbol{w}_{i\mu}^{\top} \boldsymbol{\psi}_{\mu})}{\boldsymbol{w}_{i\alpha}^{\top} \boldsymbol{\psi}_{\alpha}} \, [\exp(\boldsymbol{w}_{i\alpha}^{\top} \boldsymbol{\psi}_{\alpha} t_{i}) - 1]\right)}{1 - (1- \Upsilon) \exp\left(-\dfrac{\exp(\boldsymbol{w}_{i\mu}^{\top} \boldsymbol{\psi}_{\mu})}{\boldsymbol{w}_{i\alpha}^{\top} \boldsymbol{\psi}_{\alpha}} \, [\exp(\boldsymbol{w}_{i\alpha}^{\top} \boldsymbol{\psi}_{\alpha} t_{i}) - 1]\right)}\right\}.
    \label{eq:like_MO_Gompertz}
\end{align}

Under the Bayesian paradigm, the state of knowledge or uncertainty about the model parameters $\boldsymbol{\vartheta}$ is represented through a prior distribution $\pi(\boldsymbol{\vartheta})$. The prior distribution expresses the existing information or assumptions about the underlying scientific model before observing the data. When vague or weakly informative priors are adopted, it is commonly argued that the data tend to dominate the inference, so that the analysis is driven primarily by the likelihood function. The likelihood function $\pi(\mathcal{D} \mid \boldsymbol{\vartheta})$ characterizes the contribution of the observed data $\mathcal{D}$ given the model parameters and summarizes the information provided by the data. Bayesian inference is obtained by combining the prior distribution with the likelihood, resulting in the posterior distribution. The posterior distribution of $\boldsymbol{\vartheta}$ is given by Bayes’ theorem,

\begin{equation*}
    \pi(\boldsymbol{\vartheta} \mid \mathcal{D}) = \dfrac{ \pi(\mathcal{D} \mid \boldsymbol{\vartheta}) \times \pi(\boldsymbol{\vartheta})}{\pi(\mathcal{D})},
\end{equation*}

\noindent where the quantity $\pi(\mathcal{D}) = \int_{\boldsymbol{\vartheta}} \pi(\mathcal{D} \mid \boldsymbol{\vartheta}) \times \pi(\boldsymbol{\vartheta}) \, d\boldsymbol{\vartheta}$ is the distribution of the observed data marginalized of model's parameters, acting as a normalizing constant. Most complex Bayesian analyses do not admit a closed-form expression for the posterior distribution due to the intractable integrals involved \cite{gelman1995bayesian}. Consequently, simulation-based Monte Carlo methods have become the standard approach for generating samples from the posterior distribution $\pi(\boldsymbol{\vartheta} \mid \mathcal{D})$. These samples are then used to obtain point estimates, such as the posterior mean, median, or mode, as well as interval estimates, including credible intervals and interquartile ranges. Among the most widely used Markov chain Monte Carlo (MCMC) techniques are the Metropolis–Hastings algorithm and Gibbs sampling. More recently, the state-of-the-art Hamiltonian Monte Carlo (HMC) algorithm has gained considerable attention due to its ability to generate more efficient proposals, leading to improved exploration of the posterior distribution and faster convergence. 

For Bayesian inference, we assume independence among the elements of the vector $\boldsymbol{\vartheta}$,
\begin{equation*}
    \pi(\boldsymbol{\vartheta}) = \pi(\Upsilon) \prod_{k_{\alpha}=0}^{q_{\alpha}} \pi(\psi_{\alpha k _{1}}) \prod_{k_{\mu}=0}^{q_{\mu}} \pi(\psi_{\alpha k _{2}}),
\end{equation*}

\noindent and the following prior specifications,
$$
\psi_{\alpha k_{\alpha}} \sim \mathcal{N}(m_{\alpha k_{\alpha}}, s_{\alpha k_{\alpha}}^{2}), \, k_{\alpha} = 0, 1, 2, \dots, q_{\alpha}, 
$$
$$
\psi_{\mu k_{\mu}} \sim \mathcal{N}(m_{\mu k_{\mu}}, s_{\mu k_{\mu}}^{2}), \, k_{\mu} = 0, 1, 2, \dots, q_{\mu},
$$
$$
\Upsilon \sim \mathcal{U}(a_{\Upsilon}, b_{\Upsilon}),
$$

\noindent where $(m_{\alpha k_{\alpha}}, s_{\alpha  k_{\alpha}}),  k_{\alpha} = 0, 1, 2, \dots, q_{\alpha}$, $(m_{\mu  k_{\mu}}, s_{\mu  k_{\mu}}), k_{\mu} = 0, 1, 2, \dots, q_{\mu}$, $a_{\Upsilon}$ and $b_{\Upsilon}$ are known prior hyperparameters. The posterior distribution is proportional to


\begin{align}
\label{eq:posterior}
    \pi(\boldsymbol{\vartheta} \mid \mathcal{D}) &\propto \prod_{k_{\alpha}=0}^{q_{\alpha}} \exp\left\{\dfrac{1}{2 s_{\alpha k_{\alpha}}^{2}} (\psi_{\alpha k_{\alpha}} - m_{\alpha k_{\alpha}})^2\right\} \nonumber \\
    &\times \prod_{k_{\mu}=0}^{q_{\mu}} \exp\left\{\dfrac{1}{2 s_{\mu k_{\mu}}^{2}} (\psi_{\mu k_{\mu}} - m_{\mu k_{\mu}})^2\right\} \times \dfrac{1}{(b_{\Upsilon} - a_{\Upsilon})} \times  \pi(\mathcal{D} \mid \boldsymbol{\vartheta})
\end{align}


\noindent where $\pi(\mathcal{D} \mid \boldsymbol{\vartheta})$ is the likelihood function described in \ref{eq:like_MO_Gompertz}.

As previously mentioned, the solution of \ref{eq:posterior} is not possible to solve analytically, thus we rely on HMC algorithm to select samples from the posterior distribution efficiently. 

Hamiltonian Monte Carlo (HMC) \cite{duane1987hybrid} \cite{neal2011mcmc} ensures adequate exploration of the target density by utilizing gradient information of the target distribution to generate informed, long-range proposals, thereby enhancing sampling efficiency. This method is grounded in the principles of Hamiltonian mechanics, providing an analogy from physical dynamics for the sampling process.

The HMC algorithm conceptualizes the sampling as the evolution of a physical system. Let $\boldsymbol{\vartheta} = (\vartheta_{1}, \vartheta_{2}, \dots, \vartheta_{M})$ denote the position vector, corresponding to the model's parameters of interest. An auxiliary momentum vector $\boldsymbol{\varsigma} = (\varsigma_{1}, \varsigma_{2}, \dots, \varsigma_{M})$ is introduced, typically drawn from a zero-mean multivariate Gaussian distribution, $\boldsymbol{\varsigma} \sim \mathcal{N}(0, C_{\boldsymbol{\varsigma}})$. The total energy of the system is described by the Hamiltonian function 
$$ \mathcal{H}(\boldsymbol{\vartheta}, \boldsymbol{\varsigma}) = U(\boldsymbol{\vartheta}) + K(\boldsymbol{\varsigma}),$$
which decomposes into potential energy $U(\boldsymbol{\vartheta})$ and kinetic energy $K(\boldsymbol{\varsigma})$. In the Bayesian context, the potential energy is defined as the negative logarithm of the target posterior density $U(\boldsymbol{\vartheta}) = - \ln  \pi(\boldsymbol{\vartheta} \mid \mathcal{D})$. The kinetic energy is conventionally specified as $ K(\boldsymbol{\varsigma}) = \dfrac{\boldsymbol{\varsigma}^{\top} C_{\boldsymbol{\varsigma}}^{-1} \boldsymbol{\varsigma}}{2}$.

The system's dynamics are governed by Hamilton's equations:

\begin{equation*}
    \frac{d \vartheta_{m}}{d x^{*}} = \dfrac{d K(\boldsymbol{\varsigma})}{d \varsigma_{m}}, \quad \dfrac{d \varsigma_{m}}{d x^{*}} = -\dfrac{d U(\boldsymbol{\vartheta})}{d \vartheta_{m}}, \quad m = 1, 2, \dots, M.
\end{equation*}

\noindent where $x^{*}$ represents a time variable in the physical system. These equations preserve the Hamiltonian $\mathcal{H}$, ensuring the properties of time-reversibility and volume conservation, which are critical for maintaining the correct stationary distribution.

An analytical solution to Hamilton's equations is typically intractable. Therefore, HMC employs the leapfrog integrator, a symplectic and reversible numerical scheme. Given a step size $\varepsilon$  a single leapfrog iteration from time $x^{*}$ to $x^{*} + \varepsilon$ proceeds as follows:

\begin{enumerate}
    \item Half-step momentum update:
    \begin{equation*}
        \varsigma_{m}\left(x^{*} + \dfrac{\varepsilon}{2}\right) = \varsigma_{m}(x^{*}) - \dfrac{\varepsilon}{2} \, \dfrac{d U(\boldsymbol{\vartheta}(x^{*}))}{d \vartheta_{m}};
    \end{equation*}
    \item Full-step position update:
    \begin{equation*}
        \vartheta_{m}(x^{*} + \varepsilon) = \vartheta_{m}(x^{*}) + \varepsilon \dfrac{d K\left(\varsigma\left(x^{*} + \dfrac{\varepsilon}{2}\right)\right)}{d \vartheta_{m}};
    \end{equation*}
    \item Remaining half-step momentum update:
    \begin{equation*}
        \varsigma_{m}\left(x^{*} + \varepsilon\right) = \varsigma_{m}\left(x^{*} + \dfrac{\varepsilon}{2}\right) - \dfrac{\varepsilon}{2} \, \dfrac{d U(\boldsymbol{\vartheta}(x^{*} + \varepsilon))}{d \vartheta_{m}}.
    \end{equation*}
\end{enumerate}

This sequence is repeated for a predetermined number of steps $L$ to generate a proposed state $(\vartheta^{*}, \varsigma^{*})$ from an initial state $(\vartheta, \varsigma)$. The leapfrog integrator presents erros in discretization, HMC solve this by incorporating the Metropolis acceptance step. The proposed state $(\vartheta^{*}, \varsigma^{*})$ is accepted with probability:
\begin{equation*}
    \rho_{\text{HMC}} = \min \left\{1, \exp(-  \mathcal{H}(\boldsymbol{\vartheta}^{*}, \boldsymbol{\varsigma}^{*}) + \mathcal{H}(\boldsymbol{\vartheta}, \boldsymbol{\varsigma}))\right\}.
\end{equation*}

If rejected, the chain remains at the current state $(\vartheta, \varsigma)$. Due to the approximate energy conservation of the leapfrog method, $\rho_{\text{HMC}}$ is typically high, enabling efficient exploration.

The No-U-Turn Sampler (NUTS), introduced by \cite{hoffman2014no}, automates the selection of the trajectory length $L$ when the HMC is selecting samples. The idea is to create a criterion for terminating the simulated Hamiltonian trajectory when it begins to double back on itself (called the "U-turn") thereby avoiding inefficient, redundant exploration. The algorithm builds a binary tree of leapfrog states by recursively doubling a simulated trajectory in alternating time directions. The Hamiltonian Monte Carlo with NUTS algorithm is implemented efficiently in \texttt{rstan} package \cite{rstan}, available in \texttt{R} programming language \cite{rcoreteam}.

\section{Simulation Study}
\label{sec:Simulation Studies}

Let $F(t)$ represent the improper cumulative distribution function for the survival time associated with a specified event of interest. Suppose we aim to generate a sample of $n$ independent observations that includes failure times, censoring indicators, and covariate data, while incorporating the proportion of cure fraction $p$ under the influence of covariates.
 The scheme for generating survival data with right censoring for the defective distribution presented in this work is described in Algorithm \ref{alg:gen_data}.

\begin{algorithm}[h]
\caption{Dataset generation algorithm from defective MOGD with covariates}\label{alg:gen_data}
\begin{algorithmic}
    \State Define the values of $(\psi_{\alpha 0}$, $\psi_{\alpha 1}$, $\psi_{\alpha 2}) \in \mathbb{R}^3$,  $(\psi_{\mu 0}$, $\psi_{\mu 1}$, $\psi_{\mu 2}) \in \mathbb{R}^3$, and $\Upsilon > 0$;
    \For{$i = 1$ to $n$}
        \State Generate $w_{i\alpha1} =  w_{i\mu1} \sim \text{Bernoulli}(0.5)$, $w_{i\alpha 2} =  w_{i\mu 2} \sim \mathcal{U}(0, 1)$;
        \State Determine the individual cure rate 
        $$p_i(w_{i\alpha1},  w_{i\alpha 2}, w_{i\mu1}, w_{i\mu 2}) = \dfrac{\Upsilon \exp\left(\dfrac{\exp(\psi_{\mu 0} + w_{i\mu 1} \psi_{\mu 1} + w_{i\mu 2} \psi_{\mu 2}}{\psi_{\alpha 0} + w_{i\alpha 1} \psi_{\alpha 1} + w_{i\alpha 2} \psi_{\alpha 2}}\right)}{1 - (1 - \Upsilon) \exp\left(\dfrac{\exp(\psi_{\mu 0} + w_{i\mu 1} \psi_{\mu 1} + w_{i\mu 2} \psi_{\mu 2}}{\psi_{\alpha 0} + w_{i\alpha 1} \psi_{\alpha 1} + w_{i\alpha 2} \psi_{\alpha 2}}\right)};$$
        
        \State Generate $M_i \sim \text{Bernoulli}(1-p_i(w_{i\alpha1},  w_{i\alpha 2}, w_{i\mu1}, w_{i\mu 2}))$;
        \If{$M_i=0$}
        \State Set $t^{*}_i = \infty$;
        \ElsIf{$M_i=1$}
        \State Take $t^{*}_i$ as the root of $F(t) = u$, where $u \sim \mathcal{U}(0, 1 - p_i(w_{i\alpha1},  w_{i\alpha 2}, w_{i\mu1}, w_{i\mu 2})))$ and $F(t)$ is the CDF of the Marshall-Olkin Gompertz distribution;
        \EndIf
        \State Generate $u^{*}_{i} \sim \mathcal{U}(0,\, \text{max}(t^{*}_{i}))$, considering only the finite values of $t^{*}_{i}$; 
        \State Calculate $t_i = \text{min}(t^{*}_{i}, u^{*}_{i})$ and $\delta_i = \mathbb{I}(t^{*}_{i} \leq u^{*}_{i})$;
    \EndFor
    \State The final dataset is $\mathcal{D} = \left\{(t_i, \delta_i, w_{i\alpha1}, w_{i\alpha 2}, w_{i\mu1}, w_{i\mu 2}) : i = 1, 2, \dots, n\right\}$.
\end{algorithmic}
\end{algorithm}

We carried out a Monte Carlo simulation study considering four samples sizes ($n$ = 100,  500, 1000, and 3000). For the Marshall-Olkin Gompertz regression, we considered the model with parameters $$\alpha(\boldsymbol{w}_{i\alpha1}, \boldsymbol{w}_{i\alpha2}) = \psi_{\alpha 0} + w_{i\alpha1} \psi_{\alpha 1} + w_{i\alpha2} \psi_{\alpha 2},$$ where $\psi_{\alpha 0} = -0.4$, $\psi_{\alpha 1}=0.1$ and $\psi_{\alpha 2}=-0.1$, and $$\mu(\boldsymbol{w}_{i\mu1}, \boldsymbol{w}_{i\mu2}) = \exp(\psi_{\mu 0} + w_{i\mu1} \psi_{\mu 1} + w_{i\mu2} \psi_{\mu 2}),$$ where $\psi_{\mu 0} = -2$, $\psi_{\mu 1}=2$ and $\psi_{\mu 2}=0.2$, and $\Upsilon = 0.6$. It worth mentioning that both parameters $\alpha$ and $\mu$ can be modeled using the different sets of covariates, but we opted to model with identical information to specify our regression approach does not suffer from identifiability problems. The values of the parameters was chosen based on the point estimates in the posterior means presented in the aplication presented in Section \ref{application}, except for the parameters $\psi_{\alpha2}$ and $\psi_{\mu 2}$, due to its continuous nature, we selected arbitrary values.

For the Bayesian approach, we assumed vague prior information for the parameters, i.e, $\psi_{\alpha k_\alpha} \sim \mathcal{N}(0, 10^2)$, $k_{\alpha} = 1, 2, \dots, q_{\alpha}$, $\psi_{\mu k_{\mu}} \sim \mathcal{N}(0, 10^2)$, $k_{\mu} = 1, 2, \dots, q_{\mu}$, and $\Upsilon \sim \mathcal{U}(0, 50)$, and evaluated the bias $B(\hat{\boldsymbol{\vartheta}}) = (\boldsymbol{\vartheta} - \hat{\boldsymbol{\vartheta}})$ and the coverage probability of the 95\% credible interval considering the low-information prior specification. We generated 5000 samples from the posterior distribution, excluding the first 2500 values considered as burn-in period, using the HMC-NUTS sampler. In this sense, considering 1000 Monte Carlo replicates of posterior estimation to evaluate the asymptotic behavior of the estimation. The values of $\hat{R}$, when using two chains were close to 1 for all parameters as the literature recommends \cite{gelman1995bayesian}.

\begin{figure}[h]
    \begin{subfigure}{0.5\textwidth}
    \includegraphics[width=0.9\linewidth]{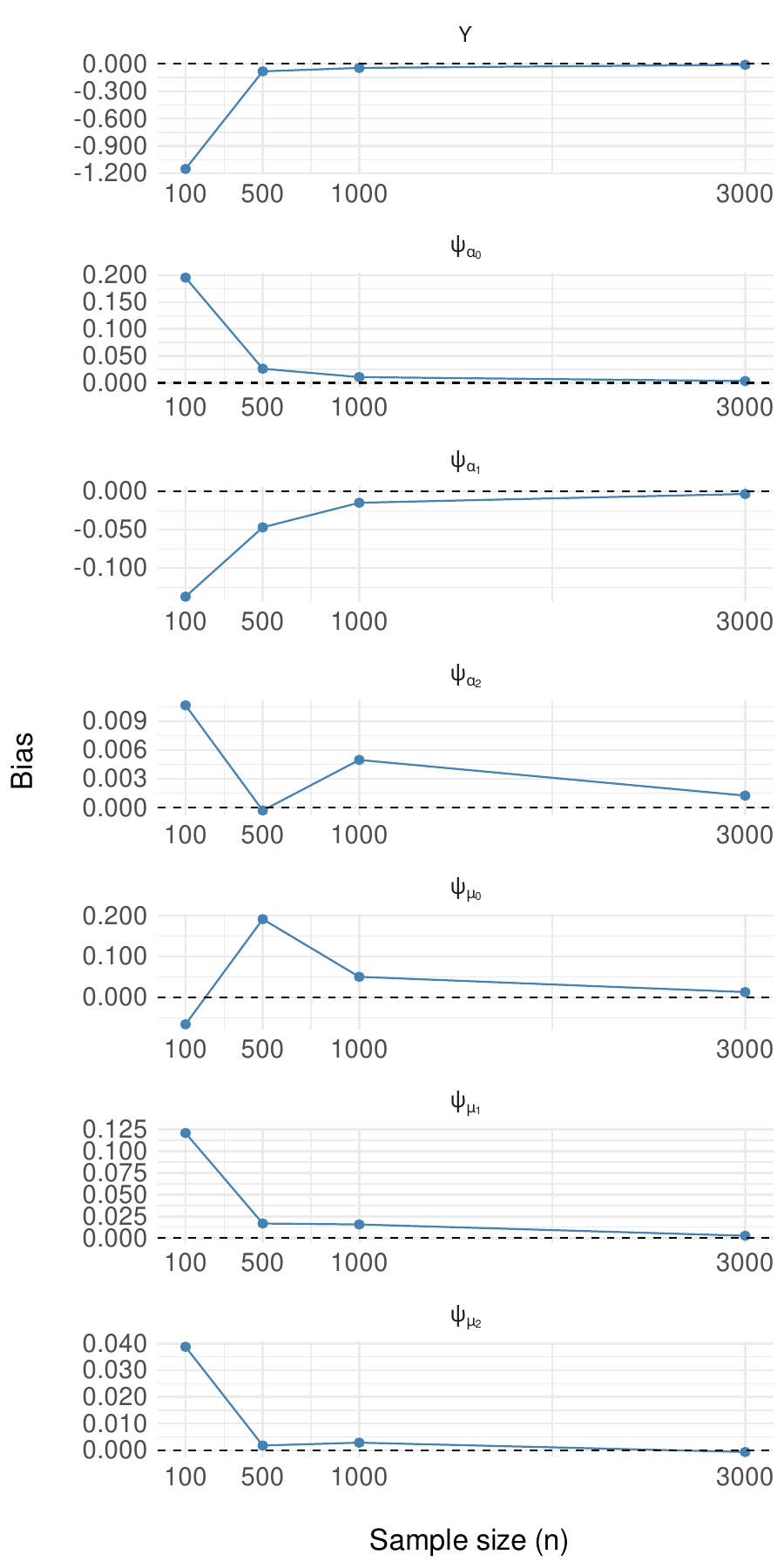} 
    \caption{}
    \label{fig:bias_a_c2}
    \end{subfigure}
    \begin{subfigure}{0.5\textwidth}
    \includegraphics[width=0.9\linewidth]{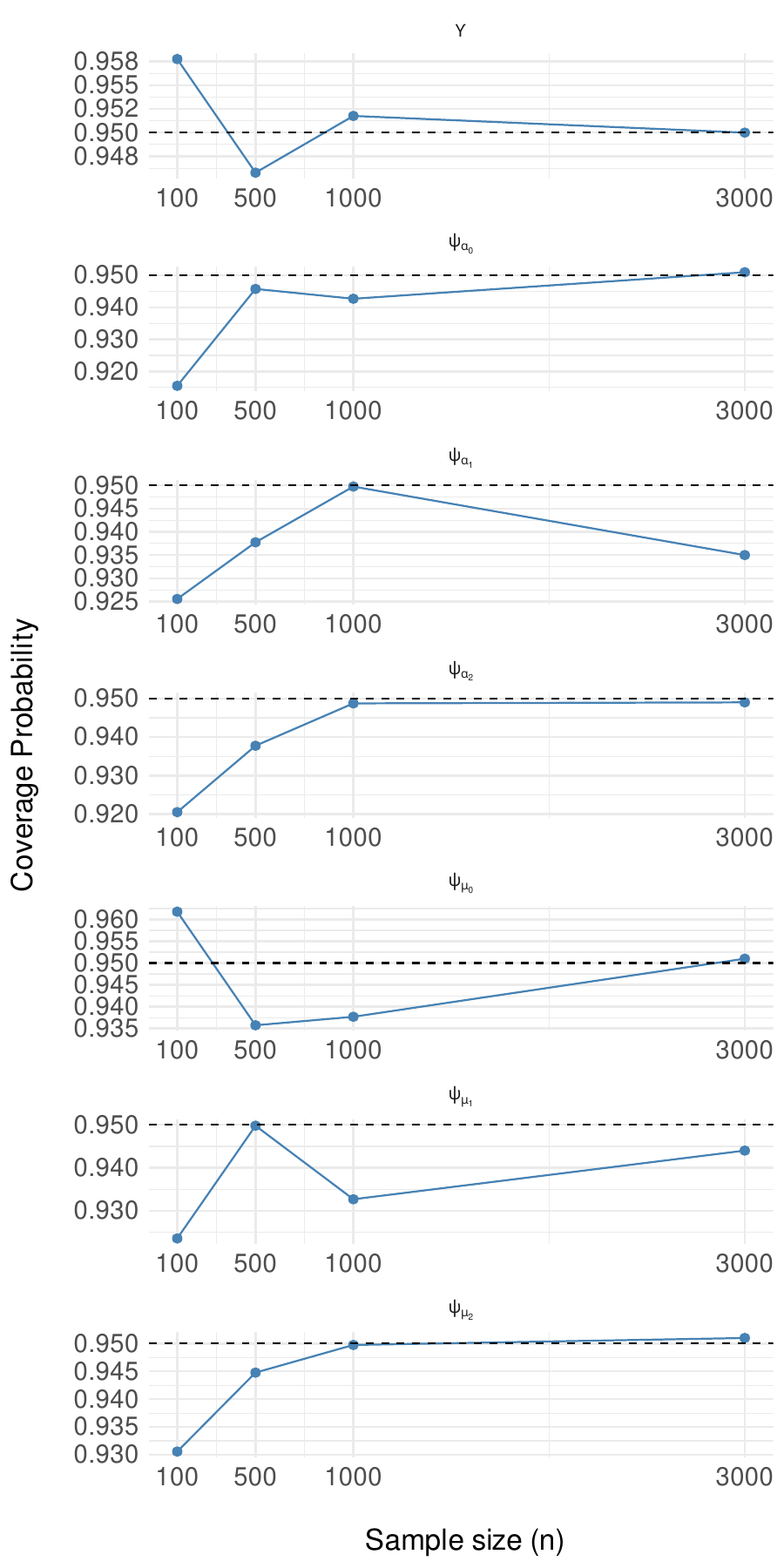}
    \caption{}
    \label{fig:cp_a_c2}
    \end{subfigure}
    \caption{Bias
     (a) and coverage probability (b) for simulated data from the Marshall-Olkin Gompertz model per sample size.}
    \label{fig:sim_results}
\end{figure}

Figure \ref{fig:sim_results} summarizes the Monte Carlo results of the bias and coverage probability for 95\% confidence intervals of $\hat{\boldsymbol{\vartheta}}$ for the different sample sizes. The simulations show the approximation of the mean of replicates to the true parameters with the decrease of variability as the sample size increases, due to the decrease of bias, showing satisfactory results for $n \geq 500$. For the sample size $n \geq 500$, the coverage probability shows values close to the nominal value of 95\%. For the cases with small samples ($n=100$), the coverage probabilities appear to be far from the nominal value for the regression parameters. In the scenario with 5000 simulated observations per replicate, comparing with the case with 100 simulated observations, we conclude that the method presents good asymptotic properties under a low informative prior information scheme.

\section{Application}
\label{application}

Testicular cancer comprises a group of malignant neoplasms (uncontrolled cell growth) that originate from different types of testicular cells, with their occurrence varying according to the group age and the cellular origin of the disease. Several cell types that constitute the testis may grow abnormally;  however, the majority of diagnosed cases correspond to germ cell tumors, which are responsible for sperm production \cite{singh2011spermatogonial}. These tumors represent the most common form of testicular neoplasm and generally affect young men between 15 and 35 years of age \cite{shanmugalingam2013global} \cite{ehrlich2015advances}.

From an epidemiological perspective, this cancer presents a relatively rare tumor type accounting for approximately 1\% of all male cancers globally \cite{purdue2005international}. The incidence and mortality of testicular cancer tend to be higher in countries with a medium Human Development Index (HDI), while high incidence rates are also observed in Nordic European countries such as Norway and Denmark \cite{znaor2014international}. In recent decades, a growing number of cases has also been reported in countries across the Latin America, including Brazil \cite{znaor2022global}. Within the national context, a higher mortality due to testicular cancer is observed in the Southeast region, particularly in the state of São Paulo, which may be associated both with a greater proportion of individuals of European ancestry and with the superior quality of epidemiological surveillance systems and cancer registries in this region \cite{soares2019testicular}.

Despite its potential severity, the prognosis of testicular cancer is highly favorable, being regarded as the most curable solid tumor. The 10-year relative survival rate exceeds 95\% \cite{verdecchia2007recent} \cite{travis2010testicular}, reflecting the efficacy of current therapeutic intervention. For localized disease, surgical treatment, typically through orchiectomy, is often sufficient to eliminate the tumor. In metastatic cases, cisplatin-based chemotherapy constitutes the main therapeutic strategy, enabling approximately 80\% of patients to achieve complete remission \cite{fossaa2006medical}. Furthermore, as diagnosis commonly occurs in young men, successful treatment often translates into an average gain of several decades of life expectancy for these patients \cite{gospodarowicz2008testicular}. However, the high cure rate is counterbalanced by the emergence of late comorbidities, frequently associated with the adverse effects of chemotherapy and other oncologic treatments \cite{schagen2008cognitive}. For this reason, long-term clinical follow-up, for at least five years after the completion of therapy, is essential to monitor recurrence and to manage long-term treatment-related complications.

The motivating dataset consists of survival information from 982 male patients diagnosed with testicular cancer between January 2015 and December 2019. The dataset was obtained from the Oncocenter Foundation of São Paulo (Fundação Oncocentro de São Paulo, FOSP), based on records from the Hospital Cancer Registry (HCR), which compiles information from residents in the state of São Paulo, Brazil. Survival time was calculated as the interval between the diagnosis date and the last recorded status, which could correspond to being cancer-free, death due to cancer, or death from other causes. The event of interest was the death of the patient in general. The dataset includes the following covariates: age at diagnosis, cancer stage (stage I, stage II, or stage III), and whether the patient received treatment (surgery, chemotherapy, etc.). For interpretation in clinical analyses, age was categorized using a relevant risk threshold for medical literature, distinguishing patients aged 35 years or older from those younger than 35. The proportion of censored observations is 0.8534, suggesting the presence of a potential cure fraction, consistent with the high censoring rate observed in the cohort and with the context reported in the literature for this type of cancer.

\begin{figure}[h]
    \centering
    \includegraphics[width=0.5\linewidth]{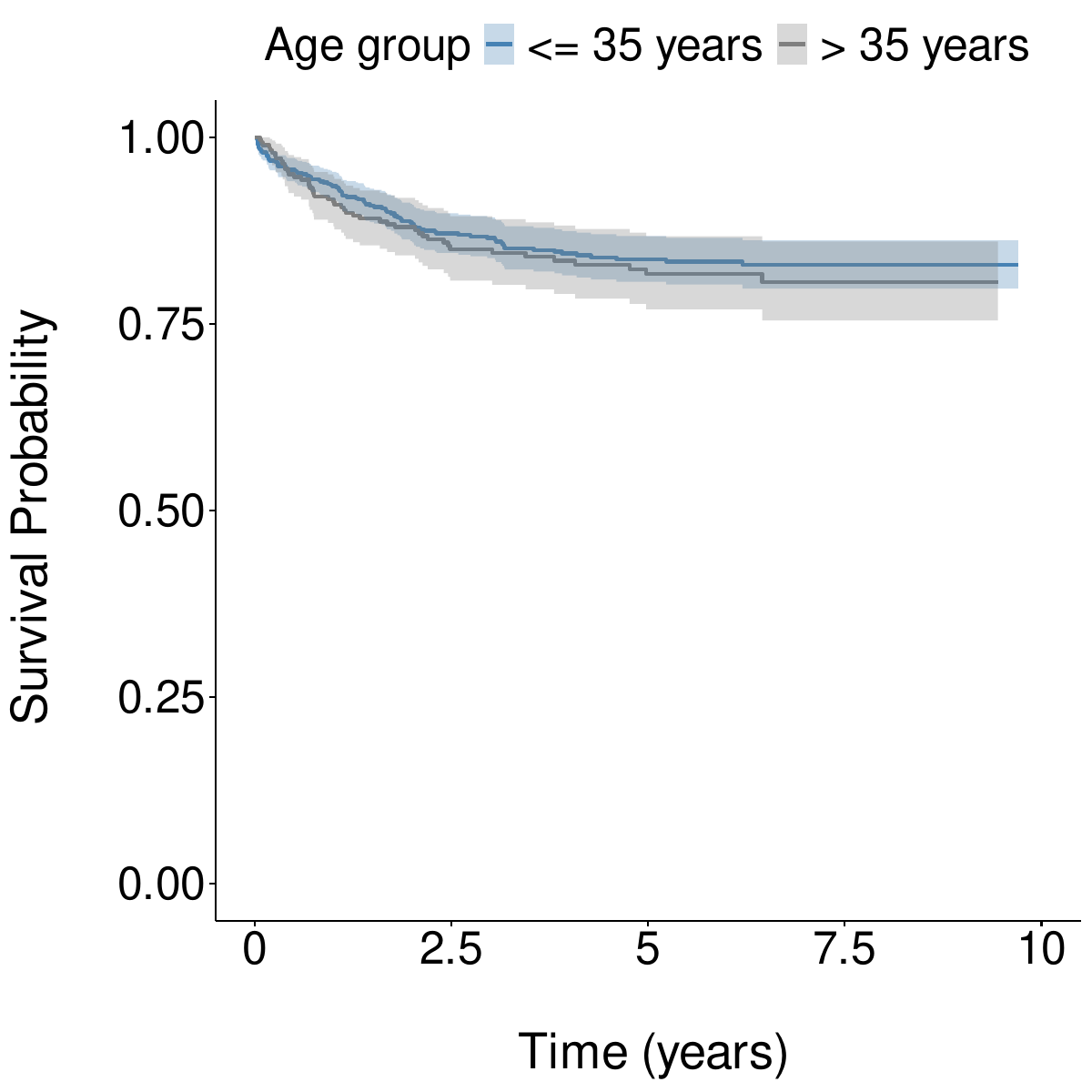}
    \caption{Kaplan--Meier curve with 95\% confidence interval of patients with testicle cancer.}
    \label{fig:kaplan_meier_curve_testicle}
\end{figure}

Figure \ref{fig:kaplan_meier_curve_testicle} presents the nonparameter Kaplan–Meier survival curve for the testicular cancer dataset, along with the corresponding 95\% confidence interval. The analysis was stratified by age group (35 years or younger vs. older than 35 years) to investigate differences in survival patterns. This threshold was chosen based on the age incidence. The non-parameter curves clearly indicate that younger patients tend to have longer survival compared with older patients. Additionally, both curves exhibit a noticeable plateau after 5 years of observation, suggesting the possible presence of long-term survivors in each age group.

\bibliographystyle{unsrt}  

For the application to real data, we incorporated the age categorization into the regression structure of the shape parameter ($\alpha$) of the defective model. In this specification, the intercept corresponds to the group of patients aged 35 years or younger, whereas the sum of the intercept and the estimated effect for individuals aged 36 years or older represents the second age group. This allows us to evaluate whether both groups exhibit evidence of a cured fraction. For the scale parameter ($\mu$), we included information on cancer stage and treatment to assess the dispersion of the data. Thus, the reference category, represented by the intercept, corresponds to male patients diagnosed at stage 1 who did not receive treatment. The regression structures for the parameters are therefore given by:
\begin{equation*}
    \alpha(\boldsymbol{w}_{i\alpha}) = \psi_{\alpha \,(\text{Intercept})} + \psi_{ \alpha (\text{Age >35})} \;w_{i (\text{Age > 35)}}, \qquad i = 1, \dots, n,
\end{equation*}
\begin{align*}
    \mu(\boldsymbol{w}_{i\mu}) &= \exp (\psi_{\mu \, (\text{Intercept})} + \psi_{ \mu \,(\text{Stage II})} \;w_{i(\text{Stage II})} \\
    & + \psi_{ \mu \,(\text{Stage III})} \;w_{i(\text{Stage III})} + \psi_{ \mu \,(\text{Treatment})} \;w_{i(\text{Treatment})}), \quad i = 1, \dots, n.
\end{align*}

The posterior estimates of the mean, standard deviation and credible interval is presented in Table \ref{tab:post_testicle}. In the analysis of both parameters associated with $\alpha$, the point estimate for the defective parameter is $-0.439$ for individuals with 35 years or younger, while for older individuals the parameter assumes $-0.439 + 0.082 = -0.357$, indicating the existence of immune individuals to the death due to testicle cancer or other clinical complications for both groups. This interpretation remains the same when we analyze the credible interval for the estimated values of $\psi_{\alpha \,(\text{Intercept})}$ and  $\psi_{\alpha \,(\text{Age >35})}$.

Due to binary information for all covariates, we can fix the parameters estimated to the parameter $\alpha$ (i.e, fix the group of analysis) and interpret the effects estimated for the parameter $\mu$ as: compared to the baseline group (men in the first stage of cancer with no treatment) represent in the intercept, if the regression coefficient is estimated as positive it decreases the probability of cure, otherwise, it increases the probability of cure. When the cancer stage advances, as expected, the probability of cure for all ages decreases, with the stage III presenting significant results in the credible interval (as it does not contains negative values). The treatment is highly important to increase the probability of cure for the testicle cancer, as it is evident in the point estimate of the posterior mean (-2.211) and in the coverage of the credible interval of the credible interval (as it does not contains positive values). 

\begin{table}[]
\centering
\normalsize
\begin{tabular}{lccc}
\hline
\textbf{Parameter} & \textbf{Mean} & \textbf{SD} & \textbf{ C.I (95\%)} \\
\hline
$\psi_{\alpha \,(\text{Intercept})}$ & -0.439 & 0.079 & (-0.604,\,-0.292) \\
$\psi_{\alpha \,(\text{Age >35})}$   &  0.082 & 0.092 & (-0.103,\, 0.258) \\
$\psi_{\mu \, (\text{Intercept})}$   & -2.258 & 1.118 & (-4.985,\,-0.601) \\
$\psi_{\mu \,(\text{Stage II})}$     &  0.535 & 0.311 & (-0.084,\, 1.129) \\
$\psi_{\mu \,(\text{Stage III})}$    &  2.330 & 0.239 & ( 1.854,\, 2.774) \\
$\psi_{\mu \,(\text{Treatment})}$    & -2.211 & 0.606 & (-3.414,\,-1.016) \\
$\Upsilon$                             &  0.568 & 0.552 & ( 0.021,\, 2.053) \\
\hline
\end{tabular}
\caption{Posterior mean, standard deviation and 95\% credible interval for testicle data.}
\label{tab:post_testicle}
\end{table}

The parameter $\Upsilon$, responsible to add flexibility in our defective regression model, contains the value 1 in the 95\% credible interval, indicating some evidence in favor of the baseline defective Gompertz distribution. For this reason, we also fitted the same regression structure under the Gompertz distribution, which corresponds to the special case $\Upsilon = 1$ of the MOGD. Model performance was then compared with that of the MOGD using Bayesian model comparison criteria, including the Deviance Information Criterion (DIC), Pareto-smoothed importance sampling leave-one-out cross-validation (PSIS-LOO), and $-2*\text{LPML}$. In all cases, lower values indicate a better-fitting model. As can be analyzed in Table \ref{tab:Bayes_metrics_chap3}, the Marshall-Olkin Gompertz model outperforms its particular case, specially in the DIC metric.

\begin{table}[h]
\centering
\normalsize
\begin{tabular}{lccc}
\toprule
\textbf{Model} & \textbf{DIC} & \textbf{PSIS-LOO} & \boldmath{$-$}{\textbf{2*LPML}} \\
\midrule
Marshall--Olkin Gompertz & 972.799 & 1004.163 & 1004.148 \\
Gompertz                 & 1003.567 & 1004.314 & 1004.313 \\
\bottomrule
\end{tabular}
\caption{Values of DIC, PSIS-LOO, and $-$2*LPML for Marshall--Olkin Gompertz and Gompertz defective regression models for testicle cancer dataset.}
\label{tab:Bayes_metrics_chap3}
\end{table}

\begin{figure}
    \centering
    \includegraphics[width=0.9\linewidth]{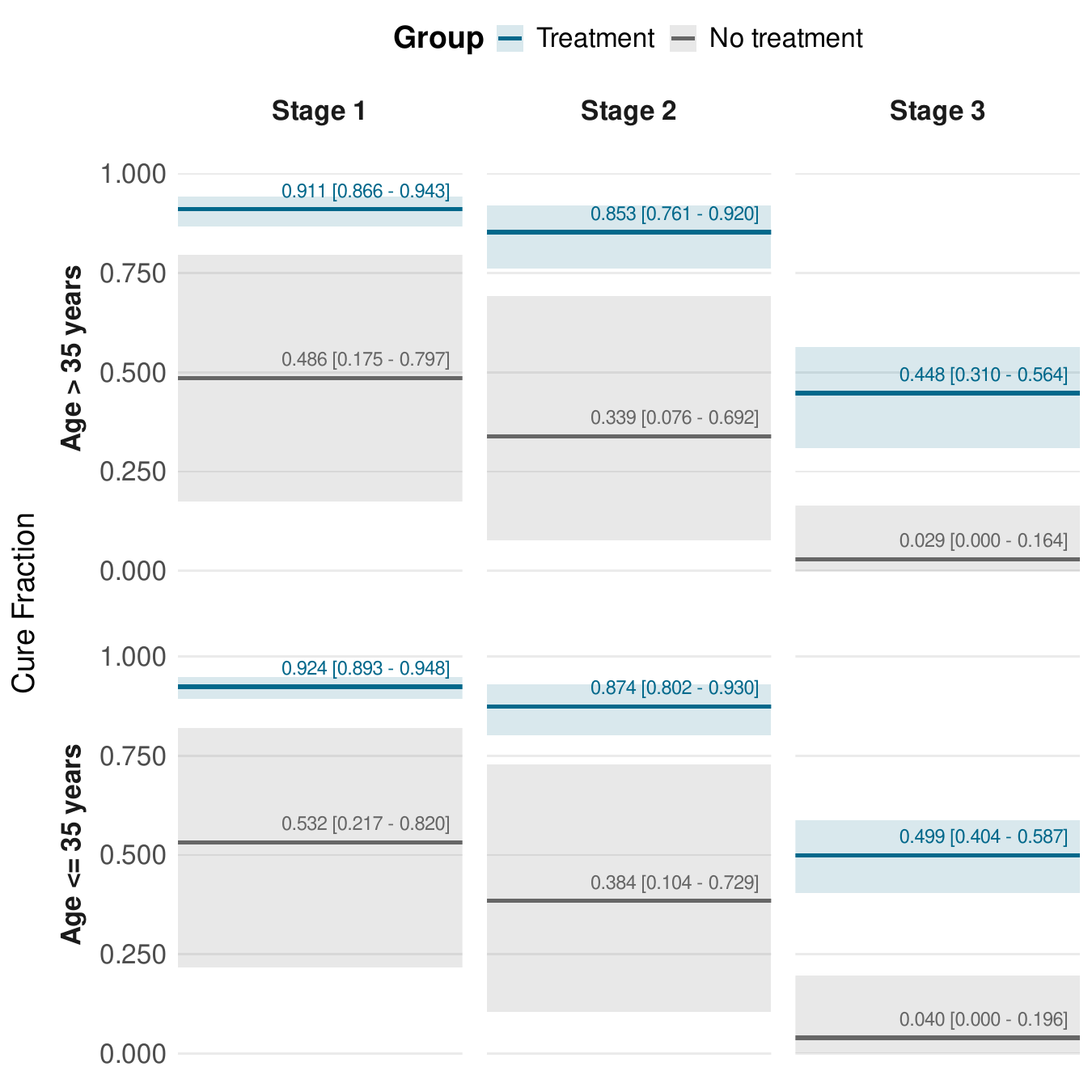}
    \caption{Posterior mean and 95\% credible interval of cure for testicle data using the MOGD.}
    \label{fig:scenarios_of_cure_app}
\end{figure}

We provide a more interpretable presentation of the fitted model by illustrating patient cure profiles using graphical summaries. Figure \ref{fig:scenarios_of_cure_app} displays posterior summaries of the cure fraction under the defective MOGD, including posterior means and credible intervals, obtained from the MCMC samples. The results are stratified by age group (35 years or younger versus older than 35 years), cancer stage (I, II, and III), and treatment status. The results highlight the importance of treatment (surgery, chemotherapy etc) in increasing survival and enhancing the probability of cure across all three cancer stages and group ages. For individuals diagnosed at stage III who did not receive any treatment, the probability of cure after after the follow-up is practically nonexistent for all individuals. When comparing the probability of long-term cure for both age groups, we observe that younger patients tend to have higher survival probabilities over time and a greater cure fraction than older patients. The probability of cure is practically not detectable for male patients at third stage of cancer and have not received any type of treatment.  

\section{Conclusions}

In this study, we present the first Bayesian approach within a regression framework for the defective MOGD, as proposed by \cite{rocha2016newtwoMO}. Our approach demonstrates to be particularly useful when aiming to identify the cure fraction stratified by groups, by introducing dichotomous covariates into the shape parameter of the distribution ($\alpha$) and modeling the scale parameter as a function of covariates. In this way, the cure probability is modeled according to two sets of covariates (which may be equal or distinct) for each individual. This formulation is particularly appealing as it helps to avoid model identifiability issues, showing satisfactory in our simulation results for sample sizes starting from 500 observations when vague priors scheme is set. Our model is also convenient to incorporate expert knowledge into modeling considering the relevance of the set of clinical covariates.

An empirical application is presented in the context of testicular cancer, which, although rare, has shown high mortality rates in cases reported in Brazil, while still exhibiting a high cure potential associated with treatment. In our study, the model identified the presence of cured individuals across different age categories, particularly among those for whom the disease has a high incidence (under 35 years old) and among those with lower incidence (over 35 years old). Within each group, the probability of cure can be further interpreted through other covariates incorporated in the scale parameter ($\mu$), where higher cancer stages were found to be detrimental to the probability of cure, while treatments such as surgery or chemotherapy were identified as beneficial factors. Additionally, we present clinical profiles to enhance the interpretation of the model, allowing readers to visualize the combination of age categories, cancer stages, and treatment status through survival curves. Credible intervals at 95\% level are also provide to quantify the uncertainty around the cure for each scenario.

Future research directions include extending the proposed framework to a competing risks context, since the dataset contains indicators of death due to cancer and death from other causes, thereby involving two distinct outcomes. Another promising line of investigation is the development of joint models based on the MOGD in this testicular cancer setting, as biomarkers play a relevant role in the clinical management of this disease \cite{dieckmann2019serum}.

\bibliography{references}  






\end{document}